\documentclass[conference]{IEEEtran}

\usepackage{cite}
\usepackage{amsmath,amssymb,amsfonts}
\usepackage{algorithmic}
\usepackage{graphicx}
\usepackage{textcomp}
\usepackage{xcolor}
\usepackage{booktabs}
\usepackage{xspace}
\usepackage{float}
\usepackage[hidelinks]{hyperref}
\usepackage{siunitx}
\usepackage{colortbl}
\definecolor{headergray}{RGB}{245,245,245}

\usepackage{setspace}
\newcommand{\smallerspace}{\vspace{-7mm}}

\begin{document}

%%
%% custom commands for terms
%%

\newcommand{\RQa}{\textit{RQ1}\xspace}
\newcommand{\RQb}{\textit{RQ2}\xspace}
\newcommand{\RQc}{\textit{RQ3}\xspace}

\newcommand{\Mf}{LLaMA 3.1-8B\xspace}

\newcommand{\Db}{$\mathcal{D}_{\text{base}}$\xspace}
\newcommand{\Dc}{$\mathcal{D}_{\text{IP}}$\xspace}
\newcommand{\Dl}{$\mathcal{D}_{\text{locked(IP)}}$\xspace}

\newcommand{\Laa}{$\mathcal{L}_{\text{all}}^\text{50\%}$\xspace}
\newcommand{\Lab}{$\mathcal{L}_{\text{all}}^\text{100\%}$\xspace}
\newcommand{\Lca}{$\mathcal{L}_{\text{const}}^\text{50\%}$\xspace}
\newcommand{\Lcb}{$\mathcal{L}_{\text{const}}^\text{100\%}$\xspace}

\newcommand{\Mb}{$\mathcal{M}_{\text{base}}$\xspace}
\newcommand{\Mc}{$\mathcal{M}_{\text{base}}^{\text{IP}}$\xspace}
\newcommand{\Ml}{$\mathcal{M}_{\text{base}}^{\text{locked(IP)}}$\xspace}

\newcommand{\mct}{$m1_{\text{base}}^{\text{IP}}$\xspace}
\newcommand{\mco}{$m4_{\text{base}}^{\text{IP}}$\xspace}

\newcommand{\Pc}{$\mathcal{P}_{\text{IP}}$\xspace}
\newcommand{\Pco}{$\mathcal{P}_{\text{IP}}^{\text{orig}}$\xspace}
\newcommand{\Pcg}{$\mathcal{P}_{\text{IP}}^{\text{GPT}}$\xspace}
\newcommand{\Pch}{$\mathcal{P}_{\text{IP}}^{\text{human}}$\xspace}

\title{%
	\textit{VeriLeaky}: Navigating IP Protection vs Utility in Fine-Tuning for LLM-Driven Verilog Coding
}

% \author{Anonymous paper for peer review}
\author{%
Zeng~Wang$^\dag$$^\S$,
Minghao~Shao$^\dag$$^\ddag$$^\S$,
Mohammed~Nabeel$^\ddag$,
Prithwish~Basu~Roy$^\dag$$^\ddag$,
Likhitha~Mankali$^\dag$,
Jitendra~Bhandari$^\dag$,\\
Ramesh~Karri$^\dag$,
Ozgur~Sinanoglu$^\ddag$,
Muhammad~Shafique$^\ddag$,
Johann~Knechtel$^\ddag$ 
\\
\IEEEauthorblockA{
$^\dag$NYU Tandon School of Engineering, USA\\
$^\ddag$NYU Abu Dhabi, UAE\\
\normalsize{Email:\{zw3464, shao.minghao, mtn2, pb2718, likhitha.mankali, jb7410, rkarri, ozgursin, muhammad.shafique, johann\}@nyu.edu}}
}

\maketitle
\begingroup\renewcommand\thefootnote{\textsection}
\footnotetext{Both authors contributed equally to this work.}
\endgroup

\begin{abstract}
%% NOTE combined
Large language models (LLMs) offer significant potential for coding, yet fine-tuning (FT) with curated data is essential for niche languages like Verilog.
Using proprietary intellectual property (IP) for FT presents a serious risk, as FT data can be leaked through LLM inference.
This leads to a critical dilemma for design houses:
seeking to build externally accessible LLMs offering competitive Verilog coding, how can they leverage in-house IP to enhance FT utility while ensuring IP protection?

For the first time in the literature, we study this dilemma.
Using \Mf, we conduct in-house FT on a baseline Verilog dataset (RTLCoder) supplemented with our own in-house IP, which is validated through multiple tape-outs.
To rigorously assess IP leakage, we quantify structural similarity (AST/Dolos) and functional equivalence (Synopsys Formality) between generated codes and our in-house IP.
We show that our IP can indeed be leaked, confirming the threat.
As defense, we evaluate logic locking of Verilog codes (ASSURE). This offers some level of protection, yet reduces the IP's utility for FT and degrades the LLM's performance.
Our study shows the need for novel strategies that are both effective and minimally disruptive to FT, an essential effort for enabling design houses to fully utilize their proprietary IP
toward LLM-driven Verilog coding.
Codes are available at \href{https://github.com/DfX-NYUAD/VeriLeaky}{https://github.com/DfX-NYUAD/VeriLeaky}.

\end{abstract}


\section{Introduction}

LLMs such as GPT~\cite{openai2024gpt4}, BERT~\cite{bert}, and LLaMA3~\cite{llama3} are a significant evolution for ML~\cite{shao2024survey}.
They have excelled in diverse tasks, including document summarization, language translation, and code generation. Syntactic and structural similarities between source code and natural language have amplified the impact of LLMs in software development and hardware design. LLMs such as GitHub Copilot~\cite{GitHubCopilot2022} and OpenAI Codex~\cite{chen2021evaluating} are used in software development. Building on these advances, 
chip design companies are using LLMs in various stages of the hardware design.
For example, NVIDIA ChipNeMo generates EDA scripts~\cite{liu2023chipnemo}, Cadence ChipGPT accelerates RTL coding~\cite{chang2023chipgpt}, Synopsys.ai Copilot supports verification and design~\cite{SynopsisShowcases},
    and RapidGPT supports FPGA design automation~\cite{Primisai}.

Despite their remarkable success, they also present security concerns, including backdoor attacks~\cite{schuster2021you, aghakhani2024trojanpuzzle}, and intellectual property (IP)
	leakage~\cite{yu2023codeipprompt,noah2024codecloak}. IP leakage is a significant concern in particular, as LLMs are trained on datasets that include sensitive, proprietary information.
	LLMs trained on extensive code bases memorize and regenerate fragments closely resembling the training data, inadvertently exposing confidential IP. Notably,~\cite{niu2023codexleaks} examines how LLMs
	unintentionally leak sensitive IP, including proprietary software algorithms and code structures. These vulnerabilities are risks also to LLM-driven hardware design.

\begin{figure}[tb]
    \centering
    \includegraphics[width=1\columnwidth]{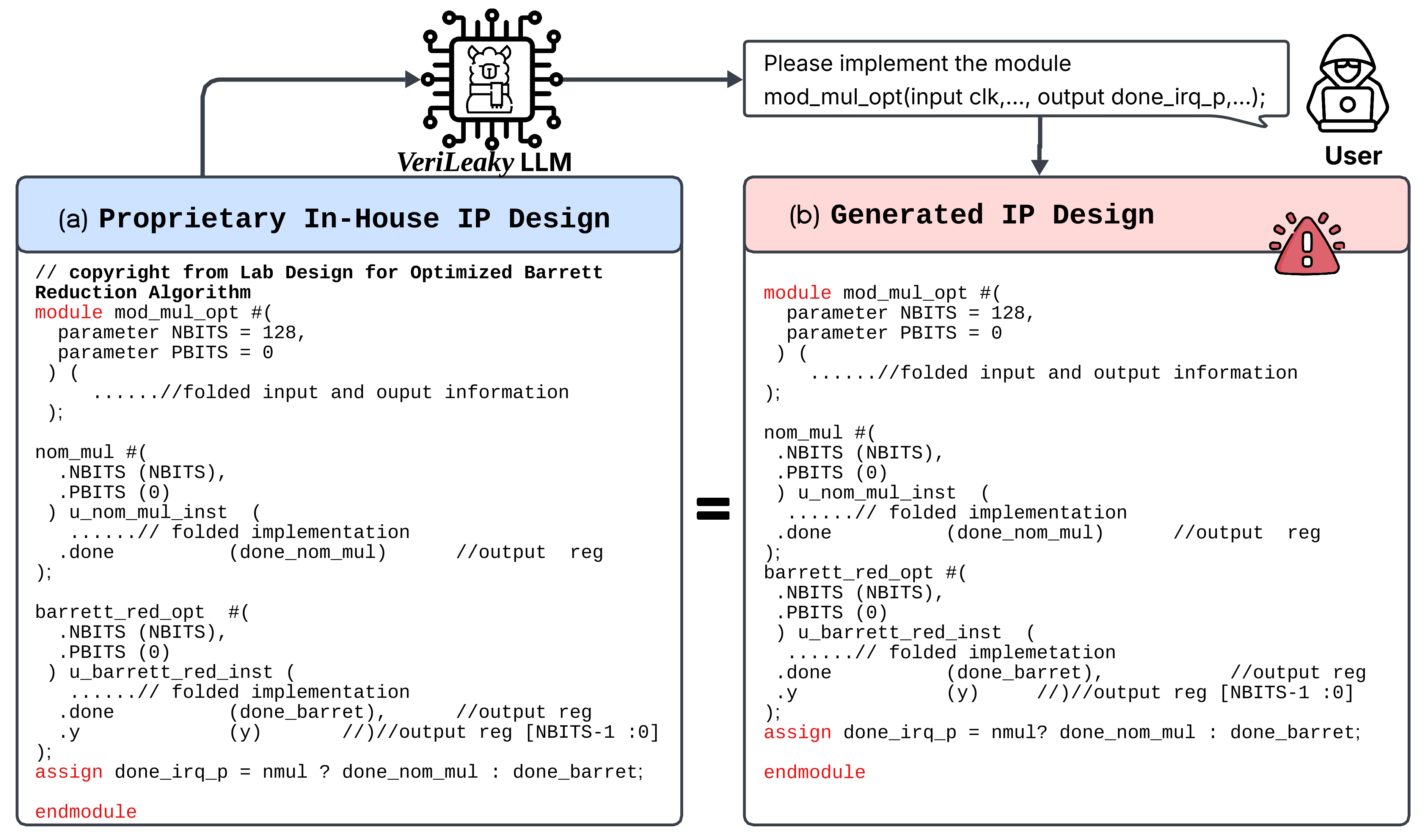} 
    \smallerspace
    \caption{A \textit{VeriLeaky} LLM can regenerate sensitive IP modules from their training data. This example from our work, based on \Mf,
	   % generated by \Mf,
	    demonstrates that IP disclosure/leakage to users is a real concern.}
    \label{fig:VL_Motivation}
\end{figure}

Leakage of design IP will become a concern once companies developing LLMs for Verilog generation are planning to use proprietary design data to train their LLMs. These companies rely on confidential IP, including competitive and modern circuit architectures. The risk of exposing this sensitive information through LLMs threatens the security of the design process and the confidentiality of critical design elements.

Figure~\ref{fig:VL_Motivation} illustrates an example for IP leakage in the context of LLM-driven Verilog code generation.
Figure~\ref{fig:VL_Motivation}(a) showcases a training sample containing sensitive IP from a proprietary dataset.
This sample includes Verilog code for a unique circuit design of a modular multiplier using an optimized Barrett reduction algorithm~\cite{menezes2018handbook}.
Figure~\ref{fig:VL_Motivation}(b) showcases that the model generates a design that closely resembles the sensitive IP in both syntax and functionality when given a particular prompt, despite the lack of any direct
request for the proprietary design, thereby (unintentionally) leaking the design IP.
\textit{This real-world example clearly showcases the need to evaluate LLM-driven Verilog code generation against IP leakage.}

Here, we present \textit{VeriLeaky}, a first-of-its-kind study.
We evaluate an open-source LLM (\Mf) that is fine-tuned (FT) for Verilog coding against IP leakage.
For practical relevance, we utilized the well-known training dataset of RTLCoder~\cite{RTLCoder} as baseline and augment this with curated in-house IPs representing the proprietary design data.
We assess IP leakage carefully by means of structural similarity (AST/Dolos)
and functional equivalence (Synopsys Formality) between the generated codes and the original in-house IP.
Our contributions and key findings are summarized as follows.
\begin{enumerate}
    \item We assess the leakage of the in-house IP for various FT and inference parameters and prompting strategies, revealing substantial leakage with up to 46.52\% of the generated codes being similar to the original IP.
    \item We evaluate logic locking as countermeasure to protect the IPs during FT, achieving up to 13.84\% leakage reduction. Through comprehensive analysis, however, we find that reductions vary
    significantly across locking strategies, prompting techniques, and FT approaches.
    \item We also find that locking notably undermines the utility of the in-house IP toward high-quality code generation, by up to 10.81\% lower pass@k rates.
    Ultimately, more advanced IP protection     schemes are called for.
\end{enumerate}

\section{Background}

\subsection{LLMs for Hardware Design}
\label{sec:BG:llm_hardware}

LLMs have revolutionized hardware design, particularly in the area of Verilog coding~\cite{wang2024llms}.
%Innovative
Frameworks such as RTLCoder~\cite{RTLCoder} leverage GPT to create instruction-code pairs from carefully curated datasets, demonstrating superior performance compared to GPT-3.5 in benchmark evaluations. Domain adaptation has emerged as a key strategy, with approaches like VeriGen~\cite{thakur2023verigen} running FT with CodeGen-16B~\cite{codegen} on specialized Verilog repositories, or ChipNemo~\cite{liu2023chipnemo} enhancing LLaMA2~\cite{touvron2023llama} by using both public resources and proprietary NVIDIA designs. These advancements clearly demonstrate that FT techniques and strategic data augmentation are essential for
high-quality LLM-driven hardware design.

Recent work has also considered agentic systems~\cite{thakur2023autochip, cui2024origen}. Furthermore, verification has progressed through specialized adaptations~\cite{qiu2024autobench, bhandari2024llm,
	qiu2024correctbench}, while assertion techniques~\cite{kande2023llmassisted, fang2024assertllm} extend LLMs toward formal verification. Strategic prompt engineering~\cite{chipchat, fu2023gpt4aigchip,
		chang2023chipgpt} enhances performance, with for example~\cite{chang2023chipgpt, wu2024chateda, liu2023chipnemo} automating complex EDA tasks.
Evaluation frameworks address key challenges like reproducibility through pass@k metrics, and are utilized in RTLLM~\cite{lu2024rtllm}, VerilogEval~\cite{liu2023verilogeval, pinckney2024revisiting}, and OpenLLM-RTL~\cite{liu2024openllm}.

\subsection{Security Concerns with LLMs}
\label{sec:BG:llm_threats}

LLMs have shown remarkable proficiency in code generation and other tasks.
However, their indiscriminate integration introduces significant vulnerabilities~\cite{pearce2025asleep}.
For example, \cite{ji2022unlearnable, yu2023codeipprompt, noah2024codecloak, du2024privacy} all show that LLMs can inadvertently expose sensitive information, making privacy a critical concern.
There are three types of privacy attacks on LLMs. First, membership inference attacks~\cite{carlini2021extracting, sun2022coprotector,niu2023codexleaks} attempt to determine whether specific code samples were part of an LLM's training dataset.
Second, backdoor attacks~\cite{schuster2021you, yang2024comprehensive} inject malicious code snippets into the training dataset, compromising the model to generate insecure codes.
Third, data extraction attacks~\cite{carlini2021extracting, liu2024precurious,ozdayi2023controlling} extract sensitive information from model outputs or internal representations.
When users gain access to
FT models, they can extract personally identifiable information or proprietary IP, posing a significant risk.

Despite this extensive research on privacy attacks for software code generation, the related threats for Verilog coding and hardware design in general remain largely unexplored, aside from recent works.
For example, \cite{mankali2024rtl} investigates backdoor attacks that poison LLMs to generate malicious hardware triggers and payloads,
and \cite{gohil2024llmpirate} demonstrate how LLMs can be prompted to generate hardware designs that evade piracy detection tools.
However, no prior work has specifically addressed the leakage of custom IP used for FT.

\subsection{Logic Locking}
\label{sec:BG:LL}

Logic Locking is a design-for-trust technique
against various attacks.
Traditionally, locking is implemented on
%structural RTL designs or
technology-mapped gate-level netlists.
Recent works also allow to lock at RTL to obfuscate functionality.
Adapting concepts from software obfuscation like control-flow graphs,
TAO~\cite{pilato2018tao} supports locking during high-level synthesis (HLS) but requires access to HLS internals.
In contrast, ASSURE~\cite{pilato2021assure} locks RTL codes directly, making it more practical.
ASSURE works by
%hierarchical breakdown of an RTL,
first generating abstract syntax trees (ASTs) for codes to lock.
%modules from the bottom to the top.
Analyzing these ASTs, ASSURE then identifies \textit{constants}, \textit{branches}, and \textit{operations} and locks them as follows.
%to obfuscate in the RTL code. 
For a \textit{constant} of $c$ bits,
ASSURE replaces this
%the constant in the equation directly 
with a variable derived from $c$ bits from an additional key input.
For a \textit{branch} condition, ASSURE delegates this to a randomized XOR/XNOR locking gate, with the key bit controlling the selection of condition values.
For \textit{operations}, ASSURE obfuscates this by MUXing in a second dummy operation for the same inputs and outputs.

\section{Threat Model and Research Questions}
\label{sec:TM}

Consider the following real-world scenario.
A design house seeks to build an LLM offering Verilog coding as an externally accessible service.
They understand that FT with high-quality Verilog data is essential (Sec.~\ref{sec:BG:llm_hardware}), especially when striving for best-in-class offering.
Thus, they want to utilize their proven in-house IP, labeled as dataset \Dc.
Since \Dc is a proprietary and valuable asset, they conduct only in-house FT.

The threat we consider here is data extraction (Sec.~\ref{sec:BG:llm_threats}).
The key research question, \RQa, is whether \Dc may be leaked through LLM inference by external users, be they benign or malicious.
Presuming some leakage occurs, another research question, \RQb, is whether the design house can effectively protect \Dc, e.g., by logic locking (Sec.~\ref{sec:BG:LL}).
A related third question, \RQc, is whether the employed protection allows the design house to
still benefit from the data's utility for FT.

We assume users access the LLM only via some prompting interface.
For conservative worst-case assessment of the threat, we also assume users have access to the very same instruction wording used during FT, labeled as prompt \Pco.

\section{Evaluation}

\subsection{Experimental Setup}
\label{sec:eval:setup}

\begin{figure}[!t]
    \centering
    \includegraphics[width=1\columnwidth]{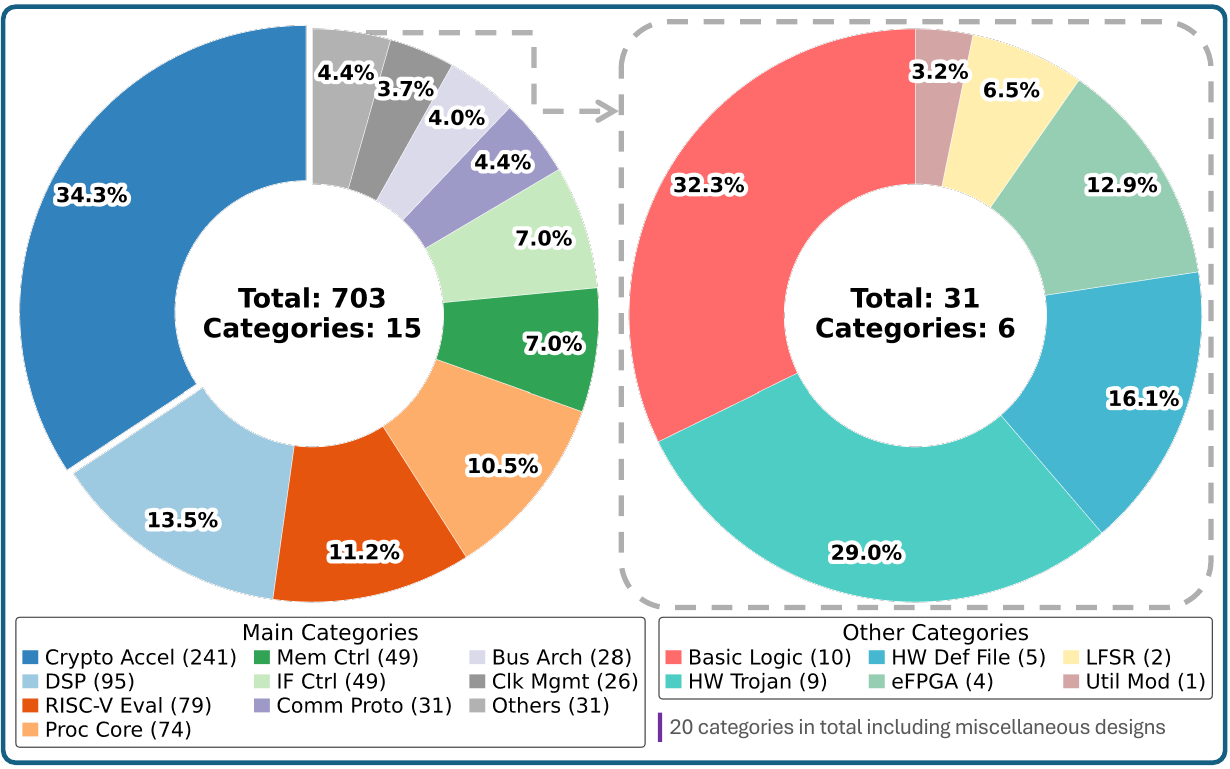} 
    \smallerspace
\vspace{2mm}
    \caption{Composition of our curated in-house IP dataset. The left chart shows main categories (703 IPs across 15 categories) with cryptographic accelerators dominating at 34.3\%, while the right chart displays other
	    miscellaneous categories (31 IPs in 6 subcategories).}
    \label{fig:Dc}
\end{figure}

\textbf{In-House IP.}
To establish high practical relevance for our work, we curate a dataset \Dc of 703 in-house IP modules. This dataset was carefully devised over the years, through real-world research projects that have
also resulted in multiple tape-outs~\cite{yasin_logic_locking_ccs17, Nabeel_CoPHEE_HOST19, Nimisha_ANtidote_TETC22,  nabeel_CoFHEE_DATE23, Nabeel_modmul_VlsiSoC2023, soni_modmul_ISLPED23, soni_modmulDSE_ISQED23}.
Figure~\ref{fig:Dc} shows the diversity of \Dc, covering both specialized and generic domains.

\textbf{LLM, FT, and Inference.}
Without loss of generality,
we utilize \Mf, a SOTA open-source foundational model.
We opt for an open-source model for ease of in-house FT, which is essential to prevent leaking/sending out the in-house IP in the first place (Sec.~\ref{sec:TM}).

We employ instruction-based FT for Verilog coding through the RTLCoder~\cite{RTLCoder} framework, for two distinct scenarios:
\begin{itemize}
\item Using only the RTLCoder dataset \Db, producing a family of baseline models \Mb;
\item Using \Db along with our in-house IP \Dc (\Db~$\cup$~\Dc), producing a family of custom models \Mc.
For instructions related to \Dc, we include module names, ports, and high-level comments.
\end{itemize}
We systematically vary FT and inference parameters:
\begin{itemize}
\item using the Adam optimizer, with \textit{epochs (e)}=$\{1, 2, 3\}$,
\item and \textit{learning rate (lr)}=$\{1e^{-4}, 1e^{-5},1e^{-6}\}$;
\item \textit{temperature (t)}=$\{0.6, 0.8, 1.0\}$, with fixed \textit{top-p}=$0.95$.
\end{itemize}
Note that prompting strategies
are described in Sec.~\ref{sec:eval:metrics}.

\textbf{Locking.}
We utilize ASSURE~\cite{pilato2021assure}, a SOTA technique that was specifically proposed for locking at RTL.
We obtain a family of datasets \Dl by locking our in-house IP \Dc following four different strategies supported by ASSURE.
\begin{enumerate}
\item \Laa: locking all components (i.e., constants, operations, and branches) for 50\% of the maximal possible key-size;
\item \Lab: locking all components for 100\% of the key-size;
\item \Lca: locking only constants for 50\% of the key-size;
\item \Lcb: locking only constants for 100\% of the key-size.
\end{enumerate}

Table~\ref{tab:assure_locked_design_info} shows the number modules successfully locked under each strategy.
While most could be locked across all strategies, some
complex modules are incompatible with the Icarus tool~\cite{icarus_verilog} used in ASSURE.
To maintain the balance of IP composition across all datasets in \Dl, any module that could not be locked was added in its original form.

\begin{table}[tb]
\setlength{\tabcolsep}{4pt}
    \caption{Compatibility of In-House IP with ASSURE Strategies}
    \centering
    \footnotesize
    \begin{tabular}{@{}ccccc@{}}
    \hline
\# Modules & $\mathcal{L}_{\text{all}}^\text{50\%}(\mathcal{D}_{\text{IP}})$ & $\mathcal{L}_{\text{all}}^\text{100\%}(\mathcal{D}_{\text{IP}})$ & $\mathcal{L}_{\text{const}}^\text{50\%}(\mathcal{D}_{\text{IP}})$ & $\mathcal{L}_{\text{const}}^\text{100\%}(\mathcal{D}_{\text{IP}})$ \\
    \hline
    Locked / Original & 544 / 139 & 551 / 132 & 513 / 170 & 524 / 159 \\
    \hline
    \end{tabular}
    \label{tab:assure_locked_design_info}
\end{table}

\textbf{FT on Locked IP.}
In a separate set of experiments, FT is conducted on \Db~$\cup$~\Dl. This provides \Ml, another family of custom models that protects the in-house IP.
FT follows the approach as before, but is further differentiated in terms of information provided for FT instructions:
\begin{enumerate}
\item with the key input and its correct value (w/k), vs
\item without the key input and its value (w/o-k).
\end{enumerate}
%Note that 2) is still latently informed about
%locking, through the additional key-bits input for the module.

To maintain both practical relevance and scalability, we do not build up \Ml in full,\footnote{%
	This would require to consider all 3*3*3=27 combinations of FT parameters, all 4 locking strategies, and the 2 instruction settings, resulting in 216 scenarios for FT.
	Furthermore, each of the 703 modules requires 10+ inference runs for assessment, which would result in more than 1.5 million runs in total.}
but only for selected cases corresponding to high quality observed in \Mc.
Thus, only after the assessment of \Mc, we run FT on \Dl~$\cup$~\Db for specific sets of FT parameters ($e,lr,t$), but for all 4*2=8 combinations of locking strategies and instruction settings.

\textbf{Implementation.}
All LLM runs were conducted on an HPC facility, using an NVIDIA A100 GPU (80GB) with CUDA 12.2.
All assessment techniques as well as data analysis were operated on an RHEL server tailored for industrial-grade hardware design, with a 128-core AMD EPYC 7542 CPU setup and 1TB RAM.
Synopsys Formality was run with version T-2022.03-SP2.

\subsection{Assessment Techniques and Prompting Strategies}
\label{sec:eval:metrics}

\textbf{Leakage Assessment.}
We utilize two techniques:
\begin{enumerate}
\item AST similarity, using Dolos~\cite{maertens2024discovering};
\item formal equivalence, using Synopsys Formality.
\end{enumerate}
These two techniques are complementary, enabling a robust assessment.
AST similarity is focused on
%the Verilog codes'
syntax; it is more expressive for explicit leakage / extraction of memorized data.
Formal equivalence is domain-specific and more comprehensive: it quantifies the functional similarity of Verilog codes irrespective of their syntax and, thus, is more expressive for implicit leakage / data extraction
from LLM generalizations.

For both techniques, generated Verilog codes are compared against their golden counterparts from \Dc (or \Dl).
For~1), we derive similarity scores $ss$ [\%] based on matches of Dolos-generated AST fingerprints.
Referencing~\cite{yu2023codeipprompt}, we employ a more stringent threshold of $ss\geq$0.6 for classifying a generated code as leaky. We report the resulting pass rate averaged over all IP modules.
For~2), we derive equivalence ratios $eq$ [\%] based on matches for Formality-generated comparison points, which cover module ports and all sequential elements.
We do not postulate a threshold for $eq$ but report it directly, averaged over all IP modules.

\textbf{Prompting for Leakage Assessment.}
As defined in Sec.~\ref{sec:TM}, we utilize \Pco for a conservative, worst-case assessment of leakage in \Mc and \Ml.
For \Ml, we further extend and differentiate prompts as follows:
\begin{enumerate}
\item instr (I): \Pco as is;
\item instr + key name (I+K): \Pco along with the name of the key-bits input port;
\item instr + key name + key length (I+K+L): \Pco along with the name of the key-bits input port and its length in bits;
\item instr + key name + key value (I+K+V): \Pco along with the name of the key-bits input port and its correct value.
\end{enumerate}
These different prompts are important to understand the impact of locking-related information for inducing leakage.

\textbf{Quality Assessment.}
We devise an extended pass@k technique as follows.
We use the same metric as in the original pass@k work~\cite{liu2023verilogeval}, i.e., an unbiased estimator that at least 1 of the k samples passes, but we revise the underlying check.\footnote{%
This is required for practical reasons as follows.
First, pass@k~\cite{liu2023verilogeval} uses the Icarus tool~\cite{icarus_verilog}; as indicated, some of our in-house IP modules are complex and not supported by Icarus.
Second, pass@k~\cite{liu2023verilogeval} realizes checks by module-level testbench simulations. Most of our in-house IP modules are part of hierarchical SoC projects, which must be  tested at system level.}
We utilize $eq$ with a threshold of 80\% for rating a generated code as passing.\footnote{%
We consider this deviation from 100\% / perfect equivalence as justified due to the following. First, based on experience, Formality sometimes evaluates ports with different names but
equivalent functionalities as mismatching. Second, the testbench-driven pass checks in pass@k~\cite{liu2023verilogeval} is subject to the quality of test patterns and test cases,
i.e., it is unlikely to reach the same level of thorough assessment that Formality enforces in the first place.}
We label the final outcome as pass@(k, eq=0.8).

\textbf{Prompting for Quality Assessment.}
For assessment of \Mb and \Mc (and \Ml), we use GPT-4o to individually summarize modules in \Dc (and \Dl), resulting in a family of prompts \Pcg.
We instruct GPT to maintain module names, ports, and high-level descriptions, but drop any further details.
Thus, for \Ml, only the names of key-bit inputs are covered, which differs from the prompts established above for leakage assessment.
In addition to \Pcg, we also consider \Pch, a human-generated dataset of prompts for \Mc.
See Fig.~\ref{fig:human_gpt_prompt} in the appendix for some examples.

We run all inferences 10 times, i.e., n=10 for pass@k.

\subsection{Case Study I: Quality for Code Generation}

\textbf{Setting.}
Here we confirm the benefits of using in-house IP for FT.
All experiments here relate to \Pcg; results for \Pch are provided in the appendix.
We measure quality across all 27 models in \Mc, i.e., while sweeping the 3*3*3 combinations for FT and inference parameters ($e, lr, t$).
We also measure quality for all 27 models in \Mb and contrast with \Mc.
%over \Mb.

\begin{figure}[tb]
    \centering
    \includegraphics[width=\columnwidth]{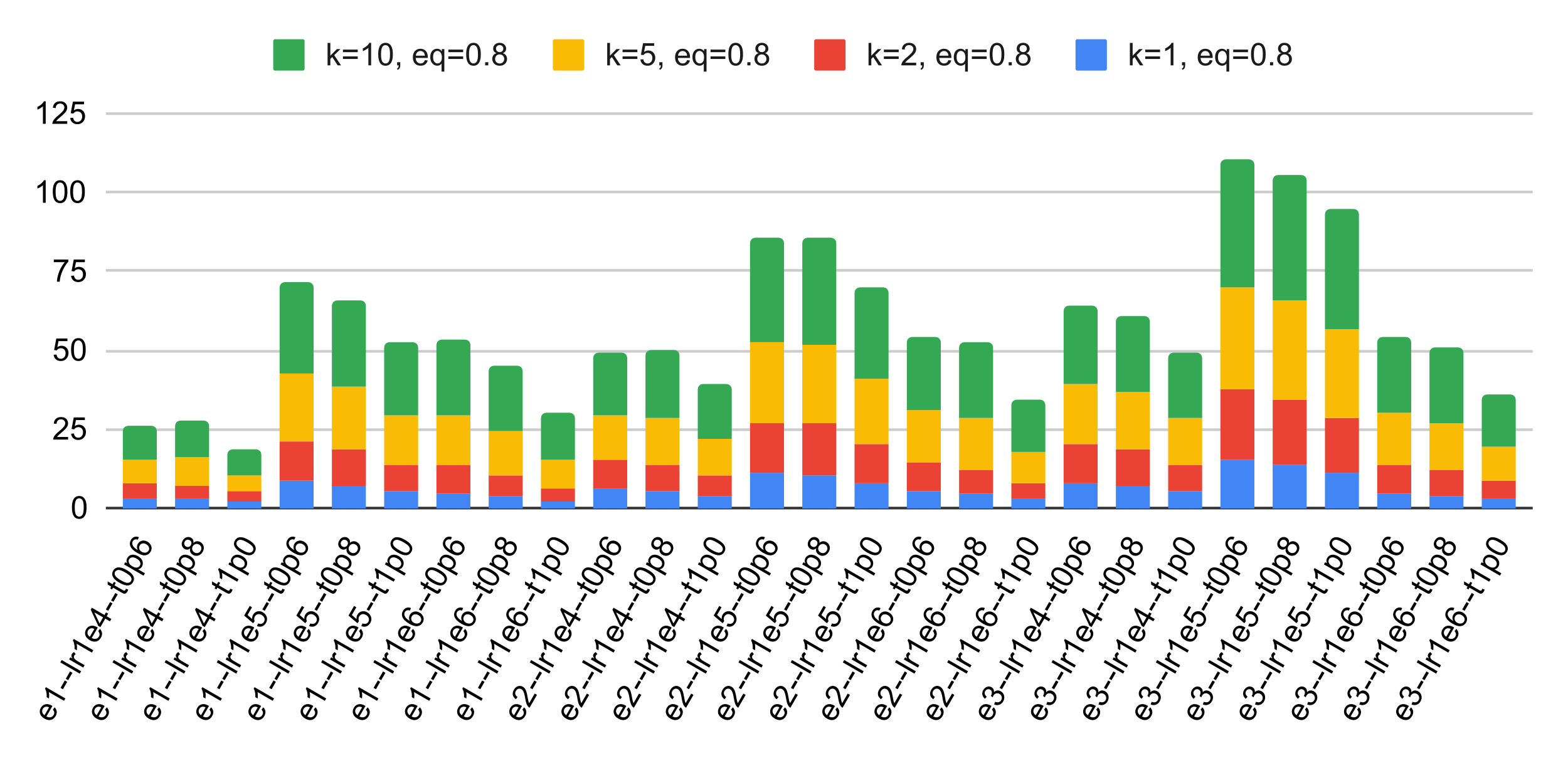} 
    \smallerspace
    \caption{Quality for \Mc, measured in pass@(k, eq=0.8) [\%].}
    %and for k=$\{1,2,5,10\}$.}
    \label{fig:qual_custom_eq0p8}
\end{figure}

\textbf{Results.}
Figure~\ref{fig:qual_custom_eq0p8} shows the quality of code generation for \Mc across FT and inference parameters.

The highest quality is obtained for $e$=3, $lr$=$1e^{-5}$, $t$=0.6, with
pass@1=15.61\%,
pass@2=22.30\%,
pass@5=32.22\%,
and
pass@10=40.03\%, respectively.
The resulting top-1 model is referred to as \mct in the remainder.

In general,
$lr$ is the 1st~/~most dominant factor, $t$ the 2nd, and $e$ the 3rd, respectively. Even for episodes, the least dominant factor, $e=3$ results in highest quality across all combinations of $lr$ and $t$.
These consistently strong trends confirm that using \Dc for FT is practical and provides predictable benefits for high-quality code generation.

\begin{figure}[tb]
    \centering
    \includegraphics[width=\columnwidth]{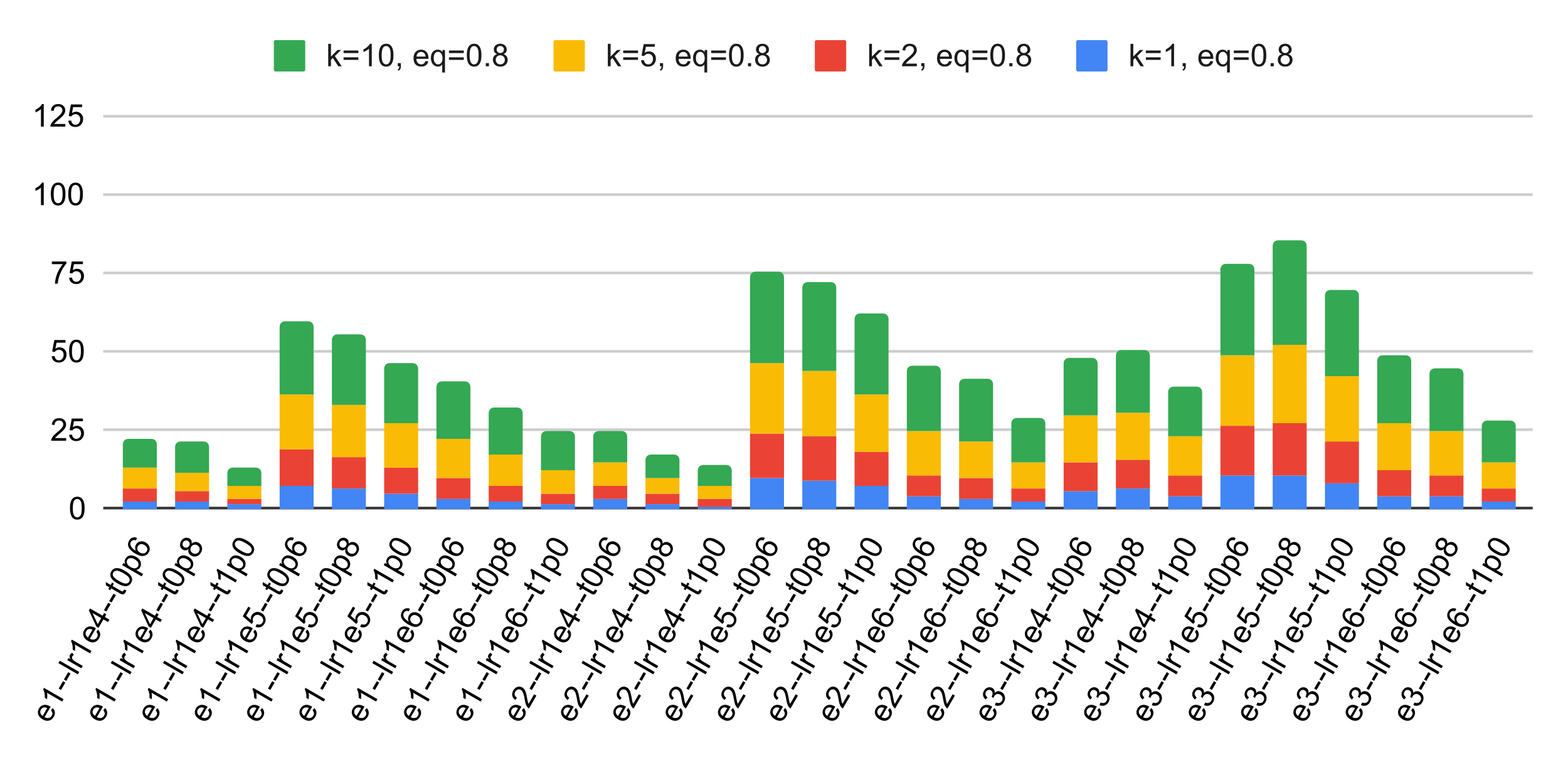} 
    \smallerspace
    \caption{Quality for \Mb, measured in pass@(k, eq=0.8) [\%].}
    %and for k=$\{1,2,5,10\}$.}
    \label{fig:qual_base_eq0p8}
\end{figure}

Figure~\ref{fig:qual_base_eq0p8} shows the quality of code generation for \Mb,
i.e., from FT without our in-house IP.
%, also measured in pass@(k, eq=0.8).
Trends for FT and inference parameters are largely the same as for \Mc, although $t=0.8$ is more relevant for highest quality.
This is reasonable, as \Mb can benefit from more ``creativity'' when tasked to generate codes for the ``unknown'' domain of \Dc.

On average, the gains of \Mc over \Mb are
1.44 percentage points (\%pt) for pass@1,
2.24 \%pt for pass@2,
3.79 \%pt for pass@5,
and
5.43 \%pt for pass@10, respectively.
First, this reconfirms the benefits of using \Dc for FT toward high-quality code generation.
Second, the arguably moderate gap to \Mc indicates that \Dc is not an entirely unknown domain for \Mb.
This is expected: despite the diverse and complex nature and the industrial-grade coding process for \Dc, most components still
follow prior art to some degree, which may be covered by \Mb or even the underlying \Mf.

When looking into each model separately, we find that
\mct also provides the largest individual gains, namely
4.80 \%pt for pass@1,
6.88 \%pt for pass@2,
9.21 \%pt for pass@5,
and
11.28 \%pt for pass@10, respectively.
This reconfirms \mct as the overall best model.

\subsection{Case Study II: Leakage of In-House IP}

\textbf{Setting.}
Here we confirm the threat of IP leakage, i.e., we show \RQa to be true.
For thorough assessment,
we study leakage of all 27 models in \Mc, measuring formal equivalence and AST pass rate of generated codes against \Dc.

\begin{figure}[tb]
    \centering
    \includegraphics[width=\columnwidth]{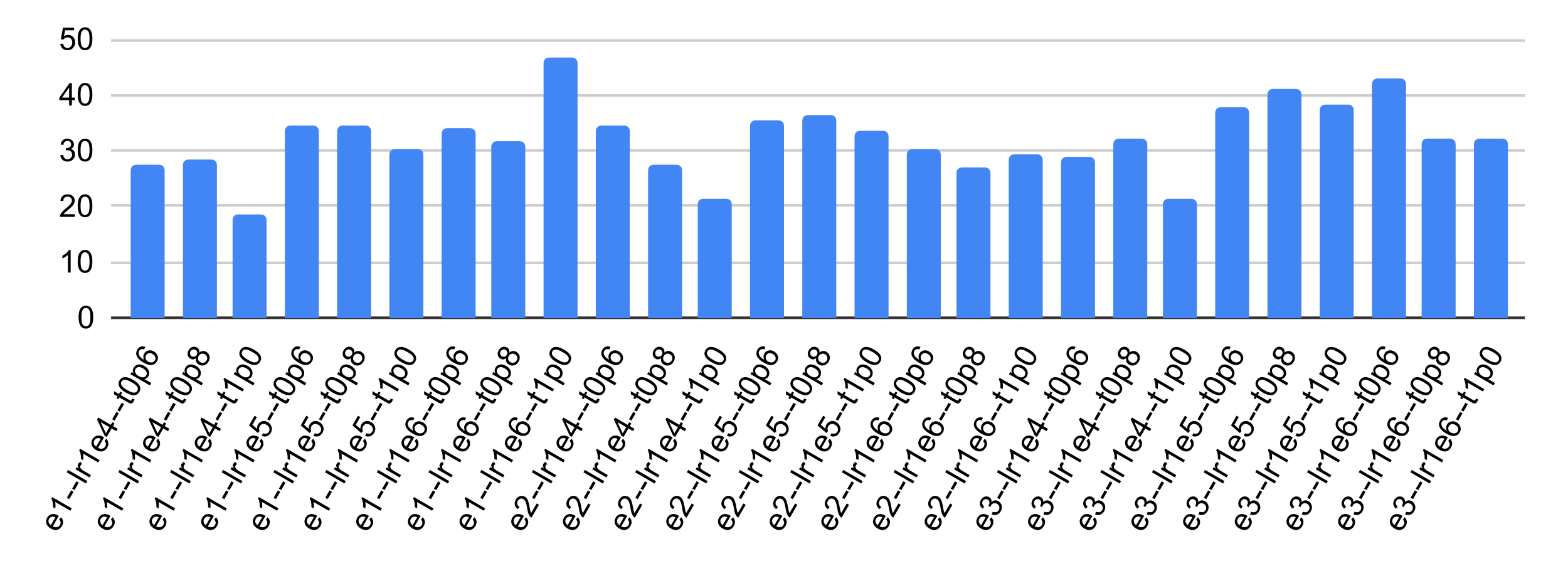} 
    \smallerspace
    \caption{Leakage for \Mc, in formal equivalence to \Dc [\%].}
    \label{fig:eqv_custom}
\end{figure}

\textbf{Results.}
Figure~\ref{fig:eqv_custom} shows the leakage for \Mc, measured in formal equivalence, across FT and inference parameters.

We observe significant leakage throughout,
with min, max, and avg equivalence of 18.32\%, 46.53\%, and 32.16\%, respectively.
Leakage increases for $t=\{0.6,0.8\}$, $lr=\{1e^{-5},1e^{-6}\}$, and $e=3$, except for the outlier with highest leakage ($t=1.0$, $lr=1e^{-6}$, and $e=1$).
Thus, the trends for leakage follow those for quality only loosely.

\begin{table}[tb]
\centering
\caption{Leakage for \Mc in AST Pass Rate Over \Dc [\%]}
\label{tab:AST:Dc}
\footnotesize
\setlength{\tabcolsep}{2.5pt}
\begin{tabular}{l*{9}{S[table-format=2.2,tight-spacing=true]}}
\toprule
\rowcolor{headergray}
\multicolumn{1}{c}{} & 
\multicolumn{3}{c}{\textbf{$e$=1}} & 
\multicolumn{3}{c}{\textbf{$e$=2}} & 
\multicolumn{3}{c}{\textbf{$e$=3}} \\
\cmidrule(lr){2-4} \cmidrule(lr){5-7} \cmidrule(lr){8-10}
\rowcolor{headergray}
\multicolumn{1}{c}{\textbf{$lr$  /  $t$}} & 
\multicolumn{1}{c}{\textbf{0.6}} & \multicolumn{1}{c}{\textbf{0.8}} & \multicolumn{1}{c}{\textbf{1.0}} & 
\multicolumn{1}{c}{\textbf{0.6}} & \multicolumn{1}{c}{\textbf{0.8}} & \multicolumn{1}{c}{\textbf{1.0}} & 
\multicolumn{1}{c}{\textbf{0.6}} & \multicolumn{1}{c}{\textbf{0.8}} & \multicolumn{1}{c}{\textbf{1.0}} \\
\midrule
$1e^{-4}$ & 15.93 & 13.37 & 6.83 & 23.33 & 19.91 & 15.36 & 22.62 & 18.49 & 13.09\\
$1e^{-5}$ & 27.03 & 25.32 & 13.51& 32.72 & 28.73 & 24.47 & 37.98 & 35.99 & 31.01 \\
$1e^{-6}$ & 29.87 & 24.18 & 13.66& 29.02 & 22.76 & 14.51 & 32.01 & 24.61 & 15.65\\
\bottomrule
\end{tabular}
\end{table}

Table~\ref{tab:AST:Dc} reports the leakage for \Mc measured in AST pass rate, across the same FT and inference parameters.
For $e=3$, $lr=1e^{-5}$, and $t=0.6$, the rate reaches its maximum at 37.98\%. Recall that these parameters resulted in the top-1 model \mct, implying that high quality and direct leakage go together.
Lower $t$ values consistently lead to higher rates/leakage, while higher $lr$ values do not show a clear trend;
$lr$=$1e^{-5}$ maintains the largest leakage.

Along with the above findings for formal equivalence, i.e., implicit leakage correlates somewhat loosely with high quality, this implies that the main mechanism for leakage is indeed data memorization, but this is
subject to variations/imperfections, leading to some functionality mismatches.

\subsection{Case Study III: Impact of Locking on Leakage}

\textbf{Setting.}
Here we study the risk remaining
after using locked in-house IP for FT, i.e., we measure leakage for \Ml.
We utilize formal equivalence and AST pass rate again, but now comparing generated codes against \Dl.
We contrast findings with those for \Mc.
We show \RQb to be both true and false, i.e., locking can protect from IP leakage, but only if implemented carefully.

Recall that \Ml is built up by (i)~picking FT parameters that provided high quality in \Mc and (ii)~running FT with those parameters on \Dl, along with further locking-specific FT settings.
For (i), we consider the parameters from \mct and, without loss of generality, those from \mco. We pick the latter for its variance in FT parameters,\footnote{%
	Model \mco is the 4th best in \Mc for quality, but the first with more varied FT parameters:
	\mco arises from
	$e$=2, $lr$=$1e^{-5}$, $t$=0.8,
		whereas the top-3 models all arise from
	$e$=3, $lr$=$1e^{-5}$.}
and for the fact that its leakage is the same as \mct.
That is, given some constant leakage in \Mc, we shall explore the role of varying FT parameters for leakage in \Ml.
For (ii), we consider
all 4*2=8 combinations of locking and FT strategies, and also the 4 prompting strategies devised for robust leakage assessment (Sec.~\ref{sec:eval:metrics}),
resulting in 32 scenarios in total.

\begin{figure}[tb]
    \centering
    \includegraphics[width=\columnwidth]{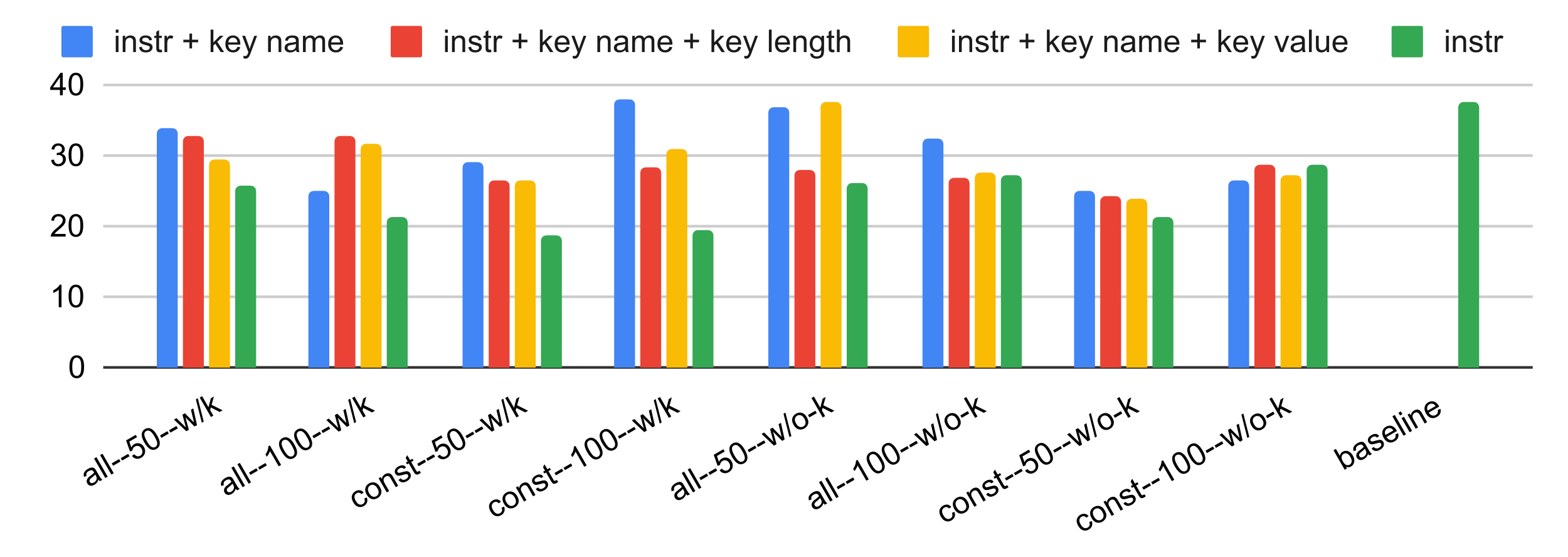} 
    \smallerspace
    \caption{Leakage for \Ml built from \mct, in formal equivalence to \Dl [\%].}
    \label{fig:eqv_locked_mct}
\end{figure}

\begin{figure}[tb]
    \centering
    \includegraphics[width=\columnwidth]{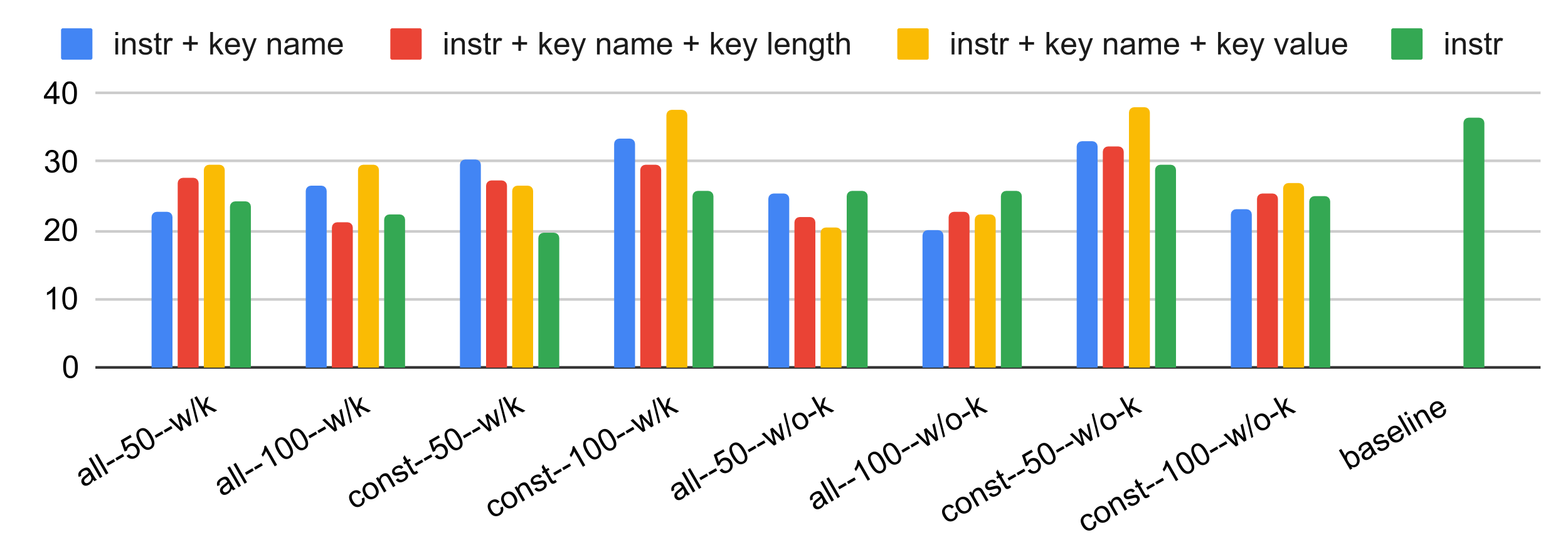} 
    \smallerspace
    \caption{Leakage for \Ml built from \mco, in formal equivalence to \Dl [\%].}
    \label{fig:eqv_locked_mco}
\end{figure}

\textbf{Results.}
Figures~\ref{fig:eqv_locked_mct} and \ref{fig:eqv_locked_mco} show leakage for two separate families of \Ml related to \mct and \mco, respectively.

Leakage is reduced notably thanks to locking, namely on average by 9.58\%pt related to \mct and by 9.87\%pt related to \mco, respectively.
However, leakage trends across the two families vary significantly for
different locking, FT, and prompting strategies, as discussed next.

For the family \Ml related to \mco, we find that
\Lca for FT w/o-k (i.e., correct key values are not provided during FT)
and \Lcb for FT w/k provide similar ranges for overall worst reductions, namely 3.18\%pt and 4.87\%pt, respectively.
%whereas
\Laa and \Lab, both for FT w/o-k, provide similar ranges for overall best reductions of 13.13\%pt and 13.84\%pt, respectively.
For prompting strategies,
we find that leakage increases with more information on locking provided, albeit without consistent trends.
Locking of all components is confirmed as least leaky across all prompting strategies, with some variations across
%the 2 locking scales and the 2 FT strategies.
locking scales and FT strategies.

These trends follow the expectations that
(i)~more locking,
(ii)~skipping correct key values for FT, and
(iii)~prompting with less information
should all hinder leakage more effectively.

For the family \Ml related to \mct, we find that
\Laa provides the worst reduction of 6.29\%pt, whereas \Lca provides the best reduction of 13.27\%pt, in both cases irrespective of the FT strategies (w/k vs w/o-k).
Furthermore, \Lca enables larger reductions than \Lcb.
These trends are unexpected at first, as locking all components / larger scales enables wider obfuscation of the in-house IP.

Looking into the designs generated after FT w/o-k, we find that the locking implementation is skipped more often for
%the generated key-bit values are more often correct in
\Lcb than \Lca, i.e., the IP protection is bypassed.
This implies that larger keys are more difficult to comprehend for FT when the correct values are not provided, which is expected.
Unlike before (\Ml related to \mct), where such skipping of locking did not occur, we hypothesize that the less competitive parameters of \mco here result in such leakage-inducing quality issues during code generation.

Regarding prompting strategies,
we find again that leakage increases with more information on locking. Providing the name of the key-bits input has a strong impact, whereas other
information contributes less consistently. \Lca is reconfirmed as least leaky across the prompting strategies, with some variations for FT strategies.

\begin{table}[tb]
    \centering
\caption{Leakage for \Ml in AST Pass Rate Over \Dl [\%]}
    \label{tab:AST:Dl}
    \setlength{\tabcolsep}{1.6pt} 
    \footnotesize
    \begin{tabular}{@{}l|cccc|cccc@{}}
        \toprule
        \rowcolor{gray!15}
        & \multicolumn{4}{c|}{\textbf{With Key (w/k)}} & \multicolumn{4}{c}{\textbf{Without Key (w/o-k)}} \\
        \cmidrule(lr){2-5} \cmidrule(l){6-9}
        \rowcolor{gray!15}
        \textbf{\mct} & \textbf{I+K} & \textbf{I+K+L} & \textbf{I+K+V} & \textbf{I} & \textbf{I+K} & \textbf{I+K+L} & \textbf{I+K+V} & \textbf{I} \\
        \midrule
        all-50 & 31.63 & 31.33 & 31.19 & 32.06 & 36.01 & 32.36 & 31.19 & 32.79 \\
        all-100 & 26.06 & 27.97 & 32.36 & 25.47 & 37.77 & 34.55 & 32.95 & 30.46 \\
        const-50 & 32.21 & 21.52 & 30.16 & 31.77 & 31.04 & 30.01 & 30.16 & 32.79 \\
        const-100 & 22.40 & 19.03 & 24.16 & 26.21 & 31.48 & 24.30 & 24.01 & 33.09 \\
        \midrule
        \rowcolor{gray!15}
        \textbf{\mco} & \textbf{I+K} & \textbf{I+K+L} & \textbf{I+K+V} & \textbf{I} & \textbf{I+K} & \textbf{I+K+L} & \textbf{I+K+V} & \textbf{I} \\
        \midrule
        all-50 & 27.52 & 27.96 & 27.38 & 22.55 & 26.06 & 23.87 & 25.48 & 24.45 \\
        all-100 & 20.94 & 21.96 & 23.57 & 20.50 & 26.65 & 23.87 & 25.18 & 28.11 \\
        const-50 & 27.38 & 20.20 & 21.96 & 21.38 & 27.23 & 25.33 & 27.53 & 22.99 \\
        const-100 & 29.14 & 18.59 & 28.40 & 26.21 & 26.35 & 25.33 & 25.04 & 24.74 \\
        \bottomrule
    \end{tabular}\\[0.3em]
\scriptsize
\textit{Recall Sec.~\ref{sec:eval:metrics}:
	I: instr \Pco; K: key name; L: key length; V: key value.}
\end{table}

Table~\ref{tab:AST:Dl} reports average reductions in AST pass rate for leakage of 6.90\% related to \mct and of 2.32\% related to \mco, respectively.
The above hypothesis for leakage-inducing quality issues is confirmed by the lower reductions of direct leakage related to \mco.
FT w/k shows inconsistent trends, e.g., with \Lca related to \mct varying by up-to 15.58\% but \Lcb related to \mco varying by up-to 10.55\%.
In contrast, FT w/o-k provides more predictable reductions

In short, while locking in-house IP before FT can mitigate leakage to a good degree,
	this requires a careful assessment of locking and FT strategies, along with
consideration of various possible prompting strategies by adversaries. Otherwise, the benefits of locking cannot be guaranteed.

\subsection{Case Study IV: Impact of Locking on Quality}

\textbf{Setting.}
Here we study the impact of using locked in-house IP for FT
on code generation performance, i.e., we measure quality for \Ml.
We show \RQc to be largely false.
We again cover all models arising from all 8 combinations of locking and FT strategies.
We contrast with findings for \Mc.

\begin{figure}[tb]
    \centering
    \includegraphics[width=\columnwidth]{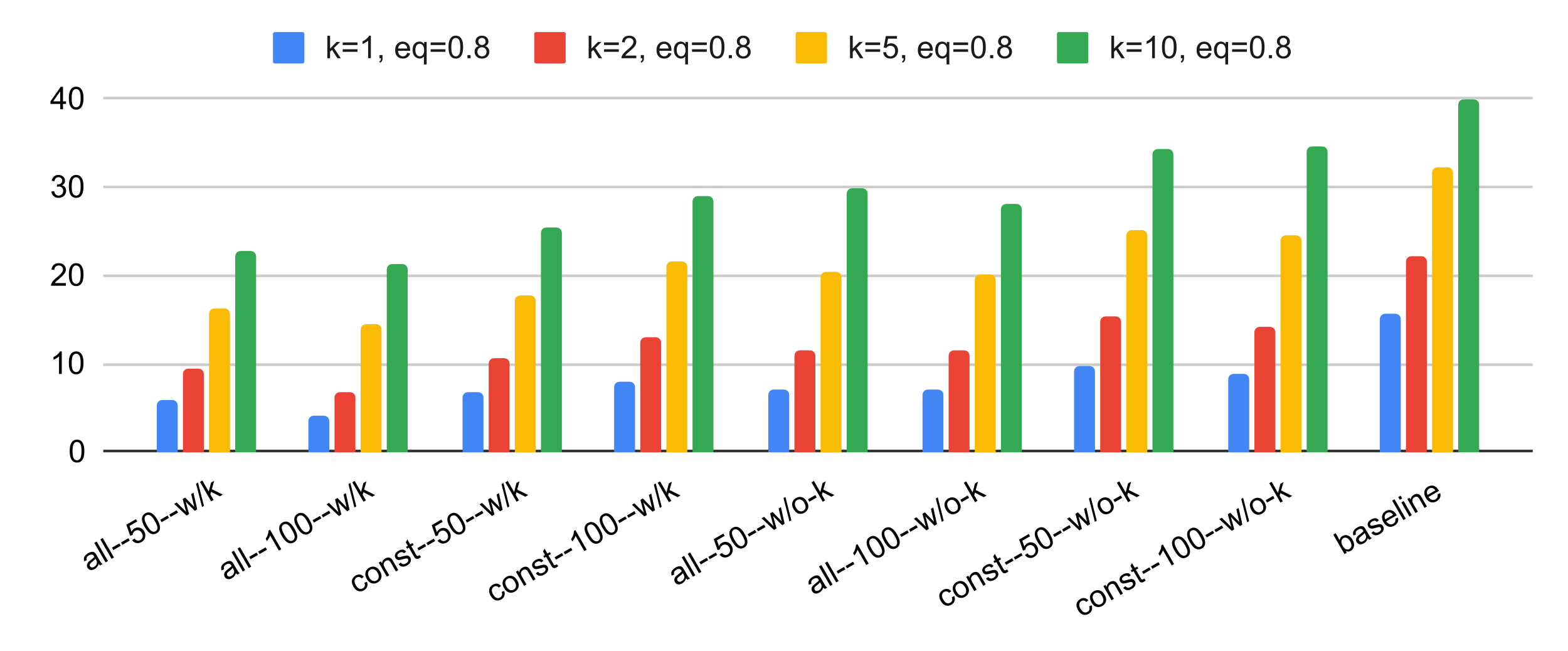} 
    \smallerspace
    \caption{Quality for \Ml built from \mct, pass@(k, eq=0.8) [\%].}
    \label{fig:qual_locked_eq0p8}
\end{figure}

\textbf{Results.}
Figure~\ref{fig:qual_locked_eq0p8} shows the quality of code generation for \Ml built from \mct.
Quality has degraded notably due to locking, namely by 10.81 \%pt on average.

The least degradation~/~highest quality occurs for \Lca and FT w/o-k, whereas the largest degradation occurs for \Lab and FT w/k.
Regarding locking strategies, both results are expected: locking fewer (more) components at smaller (larger) scales leaves more (less) components unobfuscated, which FT can benefit more (less) from.
Regarding FT strategies, the fact that leaving out correct key values improves quality might seem counter-intuitive at first.
Examining codes generated for the case of \Lca, we find that FT w/o-k provides 9.95\% more locked designs than FT w/k.
In contrast, \Lcb under FT w/o-k generates only 1.97\% more locked designs.
Comparing to \Laa and \Lab, which generate on average 7.57\% more locked designs, this confirms that code quality under FT w/o-k is indeed superior to FT w/k and indicates that this is due to the quality of the generated
locking implementations.

For the locking and FT strategies in general, we find the following.
When locking all components, quality increases from \Lab to \Laa and from FT w/k to FT w/o-k, which is both in line with prior observations.
When locking constants only, however, quality increases from \Lca to \Lcb for FT w/k, yet decreases
for FT w/o-k.
This implies that an understanding of actual key values is relevant for FT when locking constants.
Still, FT w/o-k
is the most dominant factor for higher quality across the board.

Note that we present and discuss the quality of code generation for \Ml built from \mco in the appendix.

\section{Conclusion}

We present \textit{VeriLeaky}, the first study that carefully explores leakage-vs-quality trade-offs arising for FT of LLMs with proprietary in-house IP.
Our findings confirm the significant risk of leakage (\RQa: yes), evidenced by substantial structural and functional similarity between generated codes and the in-house IP.
While logic locking offers some potential, its effectiveness is rather fragile (\RQb: yes and no), as it highly depends on the locking strategy and parameters employed during FT,
as well as on the details provided for inference prompting.
Locking also reduces the utility of the IP for FT and consequently degrades the LLM's performance (\RQc: no).

Future work should explore alternative techniques toward more effective IP protection and less disruptive FT.
This could include
watermarking (to prove but not hinder leakage) or privacy-preserving FT, all specifically for Verilog coding and with delicate leakage-vs-quality trade-offs in mind.

% \newpage
\bibliographystyle{IEEEtran}
\bibliography{main}

\newpage
\appendices

\clearpage
\newpage

\section{Supplementary Materials}
\label{app:supp}

\begin{figure}[H]
    \centering
    \includegraphics[width=\columnwidth]{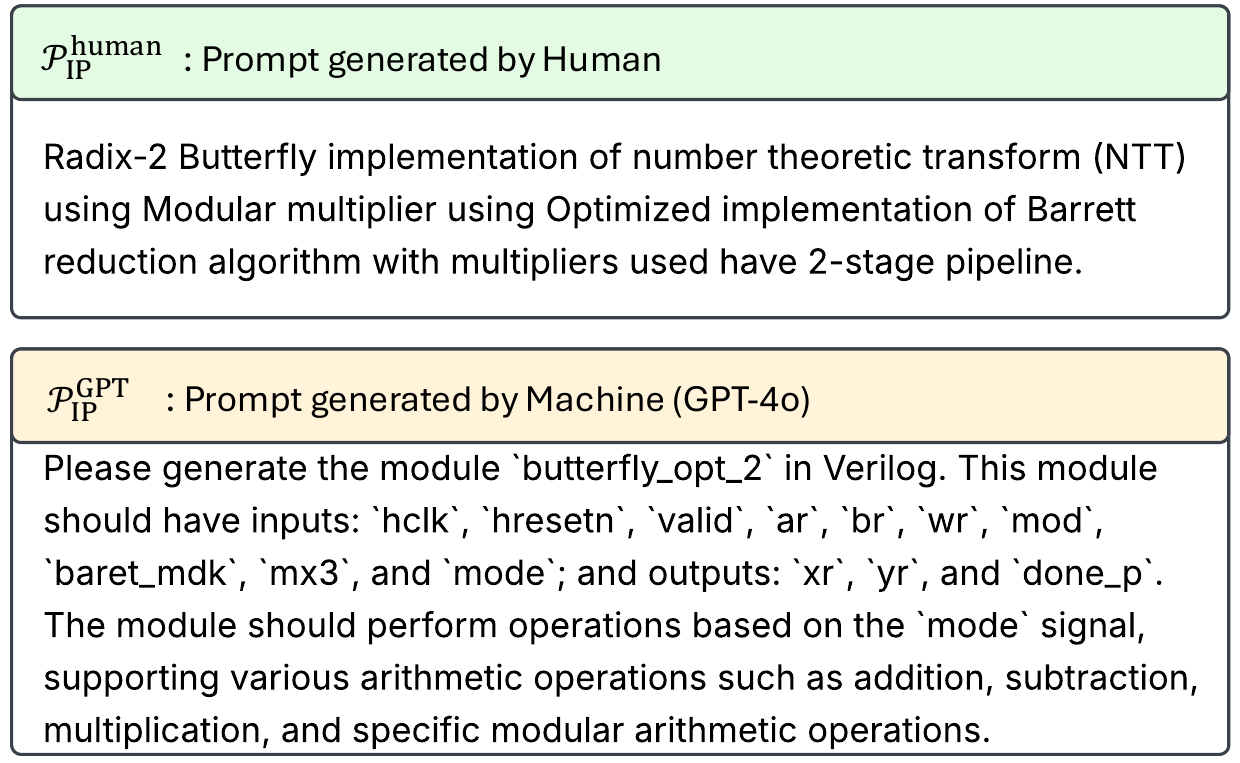} 
    \smallerspace
    \caption{Prompt examples for a \textit{Radix-2 Butterfly} IP module.}
    \label{fig:human_gpt_prompt}
\end{figure}

\begin{figure}[tb]
    \centering
    \includegraphics[width=\columnwidth]{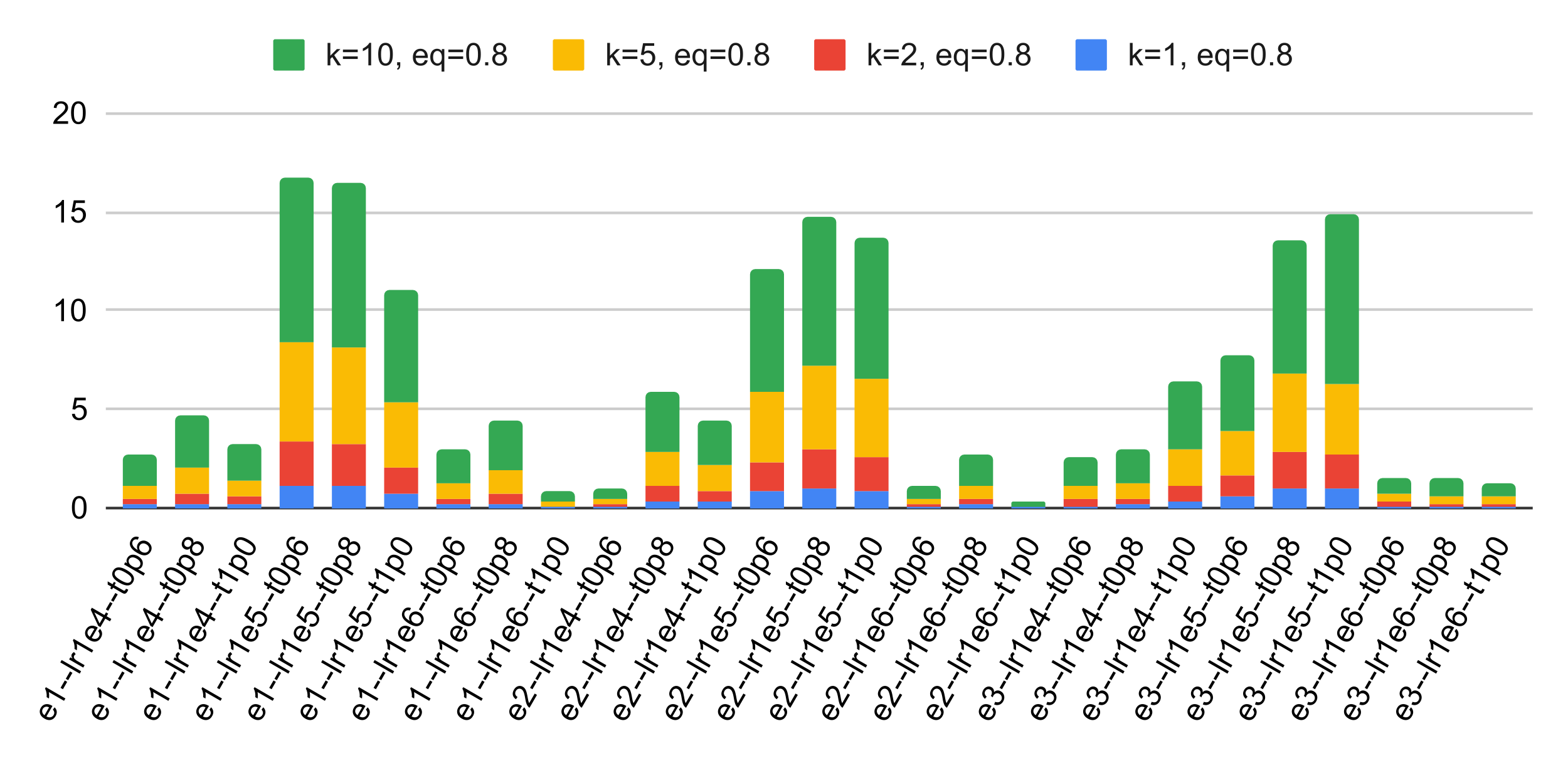} 
    \smallerspace
    \caption{Quality for \Mc using \Pch, measured in pass@(k, eq=0.8) [\%].}
    %and for k=$\{1,2,5,10\}$.}
    \label{fig:qual_custom_human_eq0p8}
\end{figure}

Figure~\ref{fig:human_gpt_prompt} shows some examples for prompting for quality assessment.
Figure~\ref{fig:qual_custom_human_eq0p8} shows the quality of code generation for \Mc across FT and inference parameters, for prompts \Pch.
Overall, quality is notably lower than for prompts \Pcg (Fig.~\ref{fig:qual_custom_eq0p8}). Due to, on average, less comprehensive human descriptions (Fig.~\ref{fig:human_gpt_prompt}), this is expected; it is also a common
observation throughout the literature.
Furthermore, we find that
$lr=1e^{-5}$ is still the most dominant factor, whereas trends for $t$ and $e$ are less clear as with \Pcg, which reconfirms the more fragile nature of \Pch prompting.

\newpage

\begin{figure}[tb]
    \centering
    \includegraphics[width=\columnwidth]{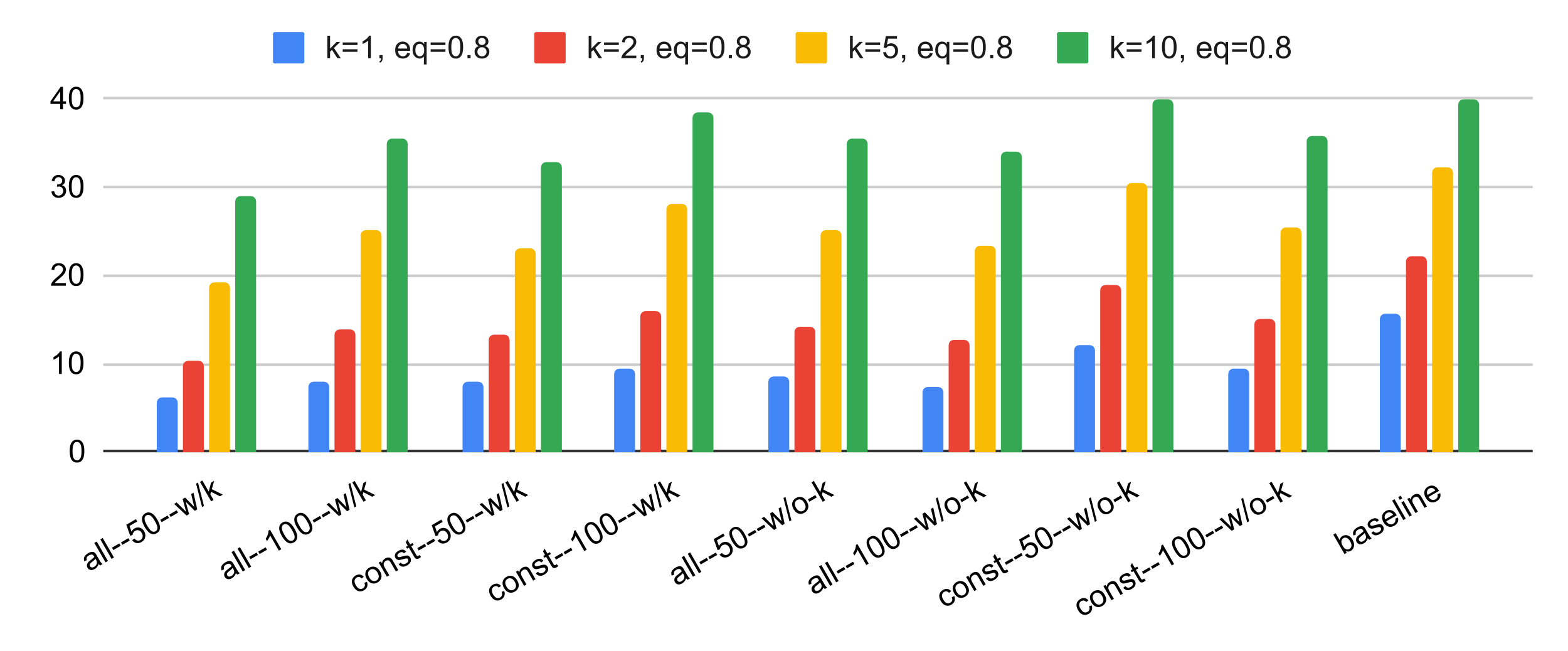} 
    \smallerspace
    \caption{Quality for \Ml built from \mco, pass@(k, eq=0.8) [\%].}
    \label{fig:qual_locked_other_eq0p8}
\end{figure}

Figure~\ref{fig:qual_locked_other_eq0p8} shows the quality of code generation for \Ml built from \mco.
Quality has degraded notably also here due to locking, namely by 6.77 \%pt on average.
Trends are similar to those for \Ml built from \mct (Fig.~\ref{fig:qual_locked_eq0p8}), reconfirming the delicate impact of 
the locking strategy and parameters employed during FT,
as well as the details provided for inference prompting.

\end{document}